\documentclass[12pt]{article}
\usepackage{epsf}
\usepackage{amsmath}
\usepackage{graphics}
\usepackage{cite}

\renewcommand{\thefootnote}{\#\arabic{footnote}}
\begin{document}

\newcommand{\gtrsim}{ \mathop{}_{\textstyle \sim}^{\textstyle >} }
\newcommand{\lesssim}{ \mathop{}_{\textstyle \sim}^{\textstyle <} }

\newcommand{\rem}[1]{{\bf #1}}

\renewcommand{\thefootnote}{\fnsymbol{footnote}}
\setcounter{footnote}{0}
\begin{titlepage}

\def\thefootnote{\fnsymbol{footnote}}

\begin{center}
\hfill August 2015\\
\vskip .5in
\bigskip
\bigskip
{\Large \bf Pre-String Theory}

\vskip .45in

{\bf Paul H. Frampton\footnote{e-mail: paul.h.frampton@gmail.com}}

\end{center}

\vskip .4in
\begin{abstract}
\noindent
In this note, I recollect a two-week period in September 1968 when
I factorized the Veneziano model using string variables in Chicago.
Professor Yoichiro Nambu went on to calculate the N-particle dual 
resonance model and then to factorize it
on an exponential degeneracy of states. That was in 1968 and the following
year 1969 he discovered the string action. I also include some other
reminiscences of Nambu who passed
away on July 5, 2105.
\end{abstract}

\begin{center}
{\bf In memory of Y. Nambu}
\end{center}

\end{titlepage}

\renewcommand{\thepage}{\arabic{page}}
\setcounter{page}{1}
\renewcommand{\thefootnote}{\#\arabic{footnote}}

\newpage

\noindent
Here my purpose is to describe early work on string theory
immediately following the appearance of the Veneziano model. In unpublished work in 1968,
a multiparticle generalization was factorized as a string theory,
and, in further unpublished work in 1969, the bosonic string action was discovered
at the University of Chicago by Professor
Yoichiro Nambu who passed away on July 5, 2015. As his postdoc, I contributed
to the very beginning of string theory in answering a question by Nambu
and factorizing the 
basic Veneziano model during September 9-23, 1968. After that I was
a spectator.

\bigskip

\noindent
I first became aware of 
the Veneziano model on
Friday September 6 1968.  I can be certain of the exact date because
it was after the ICHEP68 conference in Vienna Aug 28 - Sept 5
and before I flew to Chicago for a postdoc on Sept 7.
I did not attend the Vienna conference but my research supervisor
J.C. Taylor did. As soon as he returned to Oxford, Taylor showed
me the Veneziano model which had been widely discussed
in Vienna and was a big surprise.
For a two-particle scattering with momenta $p_1 + p_2 \rightarrow (-p_3)+(-p_4)$
the proposed model for the scattering amplitude $A(s, t)$ with 
$s=(p_1+p_2)^2$ and $t=(p_2+p_3)^2$ was
\begin{equation}
A(s, t) = \int_0^1 x^{-\alpha(s)-1} (1-x)^{-\alpha(t)-1} dx
\label{veneziano}
\end{equation}
with $\alpha(x) = \alpha(0) + \alpha^{'} x$.
It could be readily checked that this can be written as an infinite tower
of resonances in either direct ($s$) or crossed ($t$) channels. 
Remarkably Eq.(\ref{veneziano}) satisfies all FESRs in the average sense of 
DHS duality. It is manifestly $s-t$ crossing symmetric and Regge pole
behaved in both channels.
Such a simple closed solution of all of the FESRs was quite unexpected.

\bigskip

\noindent
On Monday, September 9, 1968 I reported to start work as a postdoc in the
University of Chicago
and there I met for the first time Professor Yoichiro Nambu. He was very friendly
and welcoming. He was a small man and rather quiet. At
that time his English was imperfect but readily comprehensible if one listened
carefully. He laughed and smiled a lot.

\bigskip

\noindent
When he asked what I was 
interested in, I mentioned the Veneziano model, wrote Eq.(\ref{veneziano})
on the blackboard, and discussed its properties. 
Nambu then posed to me a very interesting question
about Eq.(\ref{veneziano}): can the t-dependence be factorized as follows:
\begin{equation}
(1-x)^{-2 \alpha^{'} p_2.p_3} = F(p_2) G(p_3) ?
\label{factorization}
\end{equation}
This was a question which I could not immediately answer and neither could he.
I promised to try and that was my first assignment.
He was my first boss in my first job so I was eager to impress him. I
spent two weeks day and night wrestling with the impossible-seeming
Eq.(\ref{factorization}) and returned
two weeks later on Monday September 23 with an explicit solution.

\bigskip

\noindent
The first step was to write
\begin{equation}
(1-x) = \exp [\ln (1-x)]  
\end{equation}
and to expand the logarithm
\begin{equation}
\ln (1-x) = - \sum_1^{\infty} \frac{x^n}{n}
\end{equation}

\bigskip

\noindent
I was familiar with the Baker-Hausdorff theorem which states
\begin{equation}
e^A e^B = e^B e^A e^{[A, B]}
\end{equation}
providing that the commutator $[A, B]$ commutes with $A$ and $B$. I was familiar
also with the method of solving the quantum harmonic oscillator using operators
which satisfy $[a, a^{\dagger}] = 1$ and a ground state $|0>$ with $a|0>=0$.
To complete the solution therefore I needed an infinite number
of oscillators with what would now be called string variables

\begin{equation}
[a_{\mu}^{(m)}, a_{\nu}^{(n) \dagger}] = -g_{\mu\nu} \delta_{mn}
\label{string}
\end{equation}
with ground state $|0>$ satisfying $a_{\mu}^{(m)} |0> = 0$. 

\bigskip

\noindent
With this background and defining $F(p)$ and $G(p)$ 
\begin{equation}
F(p) = \exp \left( i \sqrt{2\alpha^{'}} p_{\mu} \sum_1^{\infty} \frac{a_{\mu}^{(n)} x^n}{\sqrt{n}} \right)
\label{F}
\end{equation}
\begin{equation}
G(p) = \exp \left( i \sqrt{2\alpha^{'}} p_{\mu} \sum_1^{\infty} \frac{a_{\mu}^{(n)\dagger}}{\sqrt{n}} \right)
\label{G}
\end{equation}
it is easily checked that the explicit solution of Eq.(\ref{factorization}) is
\begin{equation}
(1-x)^{-2 \alpha^{'} p_2.p_3} = <0| F(p_2) G(p_3) |0>
\label{solution}
\end{equation}

\bigskip

\noindent
Nambu seemed pleased and continued to work on the Veneziano
model. First he calculated the generalization from 4 to N particles
and I helped him check that his results agreed with the explicit proposals in the 
literature by Bardakci and Ruegg, and by Chan and Tsou. Nambu also
calculated the degeneracy $d(N)$ of the
level $\alpha(s)=N$ as the number of partitions of $N$ into integers,
a problem solved in 1917 by Hardy and Ramanujan, 
which for large $N$ goes as $d(N) \sim \exp (c\sqrt{N})$.

\bigskip

\noindent
All of this was well in hand before the end of the calendar year 1968 and anybody
else would have published a paper in, say,
December 1968. But Nambu had somehow
heard rumours that two other groups, one at MIT (Fubini
and Veneziano) and another at Berkeley (Bardakci and Mandelstam),
had independently discovered the exponential degeneracy. 

\bigskip

\noindent
Nambu seemed to receive adequate satisfaction just from the creativity
involved without needing to publish. An earlier example
was in 1948 in Tokyo when he calculated the one-loop
correction to the electron magnetic moment but did not publish
when he read in Time magazine that Schwinger had also done it.

\bigskip

\noindent
We waited for what seemed forever, actually until April 1969, until two papers, one
from MIT the other from Berkeley,
arrived. Both had used brute force methods such as
multinomial expansions, and had not found
Nambu's far simpler string operators.
I clearly remember Nambu's reaction was, sitting in his office
holding these two long papers with a smile,
to say: "Why did they make it so complicated?" 

\bigskip

\noindent
The 1968 results of Nambu remained unpublished. In June 1969
papers appeared by Nielsen and by Susskind saying
similar things, not knowing about Nambu's results. Priority for
string theory based 
on publications is now universally attributed to
Nambu, Nielsen and Susskind. Nambu did give a talk in June 1969
at a conference in Detroit but if, as he easily could have,
he had published a refereed paper six months earlier 
string theory would have been solely his property.

\bigskip

\noindent
In July 1969, Nambu came to my office to tell me that
the action of the string is the area of its world-sheet.
For some reason, I was so skeptical
that he bowed, apologized and went away saying he must have made a mistake.
But when I learned
that the action for a particle is the length of the world-line I
realized that it was another brilliant discovery. Just as for the string theory, however,
also in the case of the string action Nambu did not write it up until a year later
for lectures at the Copenhagen Summer School in 1970. In May 1971, 
a Japanese physicist named Goto
published it and it is generally called the Nambu-Goto action.
If he had published a refereed paper about it
in 1969, it would be the Nambu action.

\bigskip

\noindent
After leaving Chicago\footnote{During the two year period, we did successfully publish one paper
together in a festschrift for Professor Wentzel. The reference
is: Y. Nambu and P. Frampton,  Asymptotic Behavior of  Partial Widths in
the Veneziano Model, in QUANTA, A collection of scientific essays dedicated to 
Gregor Wentzel, University of Chicago Press (1970).} for CERN in 1970, we stayed in touch. 
For example in December 2002, he very kindly took me by bus and train
from Nagoya
to the nearby Meiji Mura architectural park. There was a
reconstruction of the Frank Lloyd Wright Imperial Hotel in
Tokyo where Nambu said he had stayed as a small boy before
the earthquake of 1923, so he must have
been less than three years old.
There was also
a reconstruction of the home of the Japanese writer Natsume Soseki 
whose autobiographical novel 
{\it Grass on the Wayside} Nambu greatly admired.

\bigskip

\noindent
The last time I met him was at Osaka in March 2010 when Hosotani invited me 
for a seminar and Nambu came in for the day. He had a
permanent office there but came in only very rarely. By coincidence, 
a number of students
were graduating and were delighted to have their picture taken with
him and their graduation certificates signed. 

\bigskip

\noindent
The two of us spent  a very pleasant couple of hours leafing gradually
through a book of his collected publications about which
he told many interesting anecdotes. But those publications might represent only the tip
of an iceberg. He always worked hard and wrote everything out in his neat handwriting.
In his Chicago office in 1968 he already had a lot of notebooks going back decades
all lined up on his shelves. Now there must be many more 
and I wonder what remains undiscovered there.

\end{document}